\documentclass[aps,prd,twocolumn,superscriptaddress,showpacs,byrevtex]{revtex4}

\usepackage{subfigure}

\usepackage{relsize}

\usepackage{graphicx} 
\usepackage{dcolumn}  
\usepackage{comment}

\usepackage{amssymb}
\usepackage{amsmath}
\usepackage{appendix}
\usepackage{color}

\graphicspath{{ps}}

\def\url#1{}

\begin{document}


\preprint{\vbox{ \hbox{   }
}}

\title{ \quad\\[1.0cm] Analytic Calculation of Clustered Ionization in DNA from Charged Particle Radiation 
}


\author{K. Kinoshita}\affiliation{Department of Physics, University of Cincinnati, Cincinnati, Ohio, USA} 
\author{Y. Zabarmawi}\affiliation{Department of Physics, Umm Al-Qura University, Mecca, Saudi Arabia}


\bigskip
\begin{abstract}

In previous work, we constructed a statistical framework for estimating the rate of clustered ionization from charged particle radiation in DNA.
This model is extended to include contributions from secondary electrons.
Using a simple assumption of a uniform strand-breaking probability, analytic forms for the rates of single- and double-strand break formation are derived.
The forms depend on only three parameters, to which rudimentary values are assigned.
The resulting curves are compared to results from GEANT-DNA and experimental measurements in dry DNA and found to be in moderate agreement.
This approach is readily refined and extended to more complex DNA systems and may ultimately contribute to an improved understanding of the physical origins of biological radiation effects.

\end{abstract}


\maketitle

{\renewcommand{\thefootnote}{\fnsymbol{footnote}}}
\setcounter{footnote}{0}

There is currently no universal model of ionizing radiation that correlates quantitatively with biological outcomes.
This may be somewhat surprising considering that the physical mechanisms of radiation damage have been understood for nearly a century.
It is perhaps less surprising if one considers the ability of the DNA double helix to repair itself and recover from radiation-induced breaks;
at a minimum, two ionizations must happen (nearly) simultaneously in close spatial proximity to produce an unrepairable, and thus measurable, outcome.
A unified picture of the physical origination of events leading to biological effects would likely facilitate their analysis at many scales, from DNA strand breaks to cell survival to cancer therapeutics.
We suggest that the physical model can be extended analytically in ways that are biologically relevant.
This may lead to a better understanding of the initiating events that lead to observable biological outcomes.

In previous work\cite{KK-YZ}, we presented an analytic derivation of the density of clustered ionization from charged particle radiation, starting with well-motivated simplifying assumptions about the interaction of radiation with matter.
In this picture, charged particles transfer their kinetic energy to electrons, resulting in electron vacancies (holes) in the target matter that are on the line of the radiation particle's trajectory.
The linear density of holes is proportional to the {\it stopping power} ($dE/dx$, LET), which depends on the charge and speed of the projectile and the {\it mean ionization potential}, $I_{\rm eff}$, of the target. 
$I_{\rm eff}$ was estimated for DNA via the Bragg rule\cite{Ahlen}, averaged over the four nucleotides, and found to be 83.2~eV.

The foremost motivation for understanding clustered ionization is its relevance to DNA double strand breaks, where both strands of the double helix are broken in near proximity, within $\approx$3~nm of each other along the length of the molecule.  
For this, the relevant scale for defining clustering is this distance and the width of the molecule, $\approx 2$~nm.
For simplicity, we defined a single {\it clustering scale}, $r_0$, which would be expected to be of order 2-3~nm.
Defining a cluster as two or more holes occurring on a segment of particle trajectory of length, $r_0$, we derived an expression for the density per dose of clusters as a function of LET, $\Lambda$.
In the work reported here, we address two of the limitations cited in the previous work.
First, we extend the results to include ionization from secondary electrons.
Then, by assuming a uniform probability for individual ionization events to result in strand breaks, we arrive at functional forms to describe rates of single- and double-strand breaks (SSB and DSB) and their dependence on LET.
These are compared to SSB and DSB rates calculated in simulations and measured in dry DNA, where all breaks are assumed to be from direct ionization.

{\it Secondary} ionization occurs when electrons liberated by ionization carry sufficient kinetic energy to themselves induce further ionization.
Such electrons, known as knock-on electrons or delta-rays, are considered low-LET, where the LET $\Lambda_0\approx 0.26~{\rm keV/\mu m}$ in DNA.
The expression for cluster density per dose, Eq.~4 in ~\cite{KK-YZ}, is applied to secondary radiation by substituting for $n_1$ the ionization density per dose from secondary electrons, $n_2$, and setting LET to be constant, $\Lambda=\Lambda_0=I_{\rm eff}\lambda_0$.
The total rate of clusters of $j$ holes, $\bar n^T_{{\rm clus},j}(\Lambda,r_0)$, is the sum of primary and secondary rates:
\begin{eqnarray}
\label{eqn:nclus}
\bar n^T_{{\rm clus},j}(\Lambda;r_0)&=&n_1\frac{(\Lambda r_0/I_{\rm eff})^{j-1}e^{-\Lambda r_0/I_{\rm eff}}}{j!}\\
&&+n_2\frac{(\Lambda_0 r_0/I_{\rm eff})^{j-1}e^{-\Lambda_0 r_0/I_{\rm eff}}}{j!}.
\end{eqnarray}

The secondary density $n_2$ can be estimated from $I_{\rm eff}$ and the ``$W$-value,'' the mean energy deposition per total induced ionization.
As $n_1\propto I_{\rm eff}^{-1}$ and $n_{total}=n_1+n_2 \propto W^{-1}$,
\begin{eqnarray}
n_2 = n_1\frac{I_{\rm eff}-W}{W}.
\end{eqnarray}
Although a $W$-value for DNA is not known, it is found for water and many other materials, to be of order $\sim 0.5\times I_{\rm eff}$, indicating that primary and secondary ionization contribute similarly to the total induced ionization over a range of radiation types.
We shall thus take as a first estimate that $W=0.5 I_{\rm eff}=41.6$~eV and thus $n_2=n_1=8.11\times 10^{-2}$/Gy-Mbp.\cite{KK-YZ}

\begin{figure}[h]
\includegraphics[width=8.00cm]{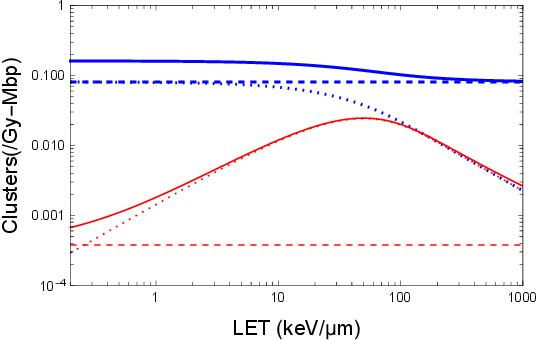}
\caption{Rates of hole clusters (/Gy-Mbp) as a function of LET (keV-$\mu$m$^{-1}$), with $r_0=3.0$~nm and $n_2=n_1$: $j\ge 1$ (thick lines) and $j\ge 2$ (thin); primary tracks (dotted), secondary tracks (dashed), and total (solid). }
\label{fig:LETdep_Calc2}
\end{figure}

Following the derivation in~\cite{KK-YZ}, the total rates for $j\ge 1$ and $j\ge 2$ are 
\begin{eqnarray}
\label{eqn:RateAll1}
\bar n^T_{{\rm clus},\ge 1}(\Lambda; r_0)&=&\frac{n_1}{\Lambda r_0/I_{\rm eff}}(1-{e^{-\Lambda r_0/I_{\rm eff}}})\nonumber\\
&&+\ \frac{n_2}{\Lambda_0 r_0/I_{\rm eff}}(1-{e^{-\Lambda_0 r_0/I_{\rm eff}}}),\label{eq:geVsLETALL}\\
\label{eqn:RateAll2}
\bar n^T_{{\rm clus},\ge 2}(\Lambda; r_0)&=&\frac{n_1}{\Lambda r_0/I_{\rm eff}}\nonumber\\
&&\times (1-{e^{-\Lambda r_0/I_{\rm eff}}}(1+\frac{\Lambda r_0}{I_{\rm eff}}))\nonumber\\
&&+\ \frac{n_2}{\Lambda_0 r_0/I_{\rm eff}}\nonumber\\
&&\times (1-{e^{-\Lambda_0 r_0/I_{\rm eff}}}(1+\frac{\Lambda_0 r_0}{I_{\rm eff}})).
\end{eqnarray}
We take as a starting point $r_0$=3~nm.
The corresponding curves are shown in Figure~\ref{fig:LETdep_Calc2}, with the primary and secondary contributions as well as the sum.

It must be assumed that not all holes created by ionizing radiation result in strand breaks.
We define a {\it strand-breaking efficiency}, $\epsilon_{\rm SB}$, to be the fraction of holes resulting in direct strand breaks and assume it is effectively uniform throughout the molecule.
We take as a first approximation the estimated fraction of electrons localized to the structural backbone of the DNA molecule, $\epsilon_{\rm SB}\approx 60\%.$\cite{KK-YZ} and examine the effect.
The densities of clustered strand breaks are found by multiplying the hole densities by $\epsilon_{\rm SB}$:
\begin{eqnarray}
n_{1,{\rm SB}}&=& \epsilon_{\rm SB} n_1\\
n_{2,{\rm SB}}&=& \epsilon_{\rm SB} n_2\\
\lambda_{\rm SB}&=&\epsilon_{\rm SB}\lambda=\epsilon_{\rm SB}\Lambda//I_{\rm eff}
\end{eqnarray}

Substituting into Eq.~4 of~\cite{KK-YZ},
the mean density per dose of clusters of $j$ strand breaks, $\bar n_{{\rm SB},j}$, is
\begin{eqnarray}
\label{eqn:nSB}
\bar n_{{\rm SB},j}&=&n_{1,{\rm SB}}\frac{(\lambda_{\rm SB}r_0)^{j-1}e^{-\lambda_{\rm SB}r_0}}{j!}\nonumber\\
&=&n_1\epsilon_{\rm SB}\frac{(\epsilon_{\rm SB}\Lambda r_0/I_{\rm eff})^{j-1}e^{-\epsilon_{\rm SB}\Lambda r_0/I_{\rm eff}}}{j!}.
\end{eqnarray}

\begin{figure}[h]
\includegraphics[width=8.00cm]{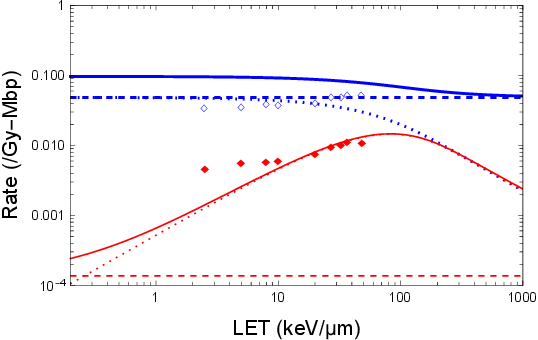}
\includegraphics[width=8.00cm]{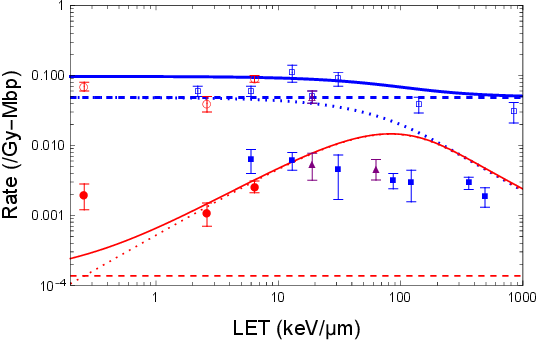}
\caption{Model curves of SSB (thick) and DSB (thin) rates as a function of LET (keV-$\mu$m$^{-1}$), from primary tracks (dotted), secondary tracks (dashed), and total (solid). 
We assume $r_0$=3.0~nm, $n_2=n_1$, and $\epsilon_{\rm SB}=0.6$.
Overlaid points are rates of direct SSB (open) and DSB (solid): (top) calculated by GEANT4-DNA Monte Carlo~\cite{Geant4}
and (bottom) measured in dry DNA, \cite{Ushigome}(squares),  \cite{Urushibara}(triangles), and  \cite{vysin}(circles).
}
\label{fig:LETdep_MC}
\end{figure}

The total SSB rate, $\bar n_{\rm SSB}(\Lambda;r_0,\epsilon_{\rm SB})$, is taken to be the sum over $j\ge 1$, including both primary and secondary ionization:
\begin{eqnarray}
\bar n_{\rm SSB}&=&\bar n^T_{{\rm SB},\ge 1}(\Lambda; r_0,\epsilon_{\rm SB})\nonumber\\
&=&\frac{n_1\epsilon_{\rm SB}}{\Lambda r_0/I_{\rm eff}}(1-{e^{-\Lambda r_0\epsilon_{\rm SB}/I_{\rm eff}}})\nonumber\\
&&+\ \frac{n_2\epsilon_{\rm SB}}{\Lambda_0 r_0/I_{\rm eff}}(1-{e^{-\Lambda_0 r_0\epsilon_{\rm SB}/I_{\rm eff}}}).
\label{eqn:RateSSB}
\end{eqnarray}

Similarly, the DSB rate, $\bar n_{\rm DSB}(\Lambda;r_0,\epsilon_{\rm SB})$, is taken to be the sum over $j\ge 2$:
\begin{eqnarray}
\bar n_{\rm DSB}&=&\bar n^T_{{\rm SB},\ge 2}(\Lambda; r_0,\epsilon_{\rm SB})\nonumber\\
&=&\frac{n_1\epsilon_{\rm SB}}{\Lambda r_0/I_{\rm eff}}\nonumber\\
&&\times (1-{e^{-\Lambda r_0\epsilon_{\rm SB}/I_{\rm eff}}}(1+\frac{\Lambda r_0\epsilon_{\rm SB}}{I_{\rm eff}}))\nonumber\\
&&+\ \frac{n_2\epsilon_{\rm SB}}{\Lambda_0 r_0/I_{\rm eff}}\nonumber\\
&&\times (1-{e^{-\Lambda_0 r_0\epsilon_{\rm SB}/I_{\rm eff}}}(1+\frac{\Lambda_0 r_0\epsilon_{\rm SB}}{I_{\rm eff}})).\label{eqn:RateDSB}
\end{eqnarray}
The corresponding curves, for primary, secondary, and total rates, are shown in Figure~\ref{fig:LETdep_MC}, under the assumptions $r_0$=3.0~nm, $n_2=n_1$, and $\epsilon_{\rm SB}=0.6$.  
Our calculation applies to {\it direct} ionization and does not include additional strand breaks that may occur from chemical interactions with the environment of the DNA.
Strand breaks in dry DNA are considered to be caused only by direct ionization.
Overlaid in Figure~\ref{fig:LETdep_MC} are rates of direct SSB and DSB rates, (top) derived from GEANT4-DNA Monte Carlo simulations~\cite{Geant4}
and (bottom) measured in dry DNA\cite{Ushigome,Urushibara,vysin}.
The curves are remarkably consistent with the simulated and measured points, given the simplicity of the model and initial values for $W$, $r_0$, and $\epsilon_{\rm SB}$, as well as unknown systematic uncertainties in data.

To summarize, we have derived an analytic form for rates of SSB and DSB production in DNA by charged particle radiation, using a simple but well-validated model of track ionization in matter that is applied to primary and secondary ionization, clustering, and strand breaking.  
Our previous calculation of clustered ionization rates has been extended to include secondary electrons.
Assuming a uniform efficiency for holes to result in strand breaks, we arrive at SSB and DSB rates as a function of LET.
Comparisons with published rates of SSB and DSB from GEANT-DNA simulation and from measurements in dry DNA over a range of LET show reasonable compatibility.
Importantly, our approach provides a conceptual and analytic foundation upon which better understanding of biological effects and links between different forms of ionizing radiation may be built in the future.


The authors thank Richard Gass and Michael Lamba for helpful discussions and for reading this manuscript.

\end{document}